\documentclass{article}
\usepackage{amssymb}

\usepackage{graphicx}
\usepackage{amsmath}

\input{tcilatex}

\begin{document}

\title{Phenomenological Description of the Yrast Lines.}
\author{\textbf{Vladimir P. Garistov} \\
{\small Institute for Nuclear Research and Nuclear Energy,Sofia, Bulgaria.}}
\maketitle

\begin{abstract}
Analysis of the collective structure of $K^{\pi }=0^{+}$\ \ states in terms
of phonon excitations for great amount of nuclei is presented . It is shown
that the rotational bands built on these states strongly depend on their
structure. The spectrum of the bands is obtained by the application of the
Nuclear Surface Oscillation /NSO/ model and assumption of ''Kassini form''
for the nuclear shape. The yrast lines\ energies can be successfully
described by the crossing of the different rotational ''$\beta $\ - bands'',
if the collective structure of these band heads is properly taken into
account.
\end{abstract}

Great amount of theoretical and experimental investigations are dedicated to
the analysis of the excited $0^{+}$\ \ states and\ the bands built on them 
\cite{1} . These investigations show that in the deformed nuclei the
dependence of the energies of the rotational bands\ on the angular momentum$%
\;L$ is qualitatively similar for the ground band and the ''$\beta $ -
bands'' constructed on each of these states. Therefore in first
approximation these bands may be considered$\;$without including the band
head structure. Nevertheless the influence of the $0^{+}$ states structure
on the rotational spectra is very important. In order to explain the
quantitative differences in the rotational bands on the different \ band
heads, as well as the energies and transition probabilities of the yrast
lines we have to take into account their collective structure. This leads to
the idea to link the structure of the $K^{\pi }=0^{+}$\ $\;$states to the
changes in behavior of rotational bands built on them. Thus\ it is
straightforward to examine some simplified models of the rotational bands of
deformed nuclei, in which the dependance on the collective structure of the 0%
$^{+}$ states is considered. Further we can explain the yrast line behavior
and structure as a result of the band crossings of these rotational bands.

\ For this purpose in this paper we first study the low-energy \ 0$^{+\;}\;$%
spectra of a large amount of the even-even nuclei from the heavier nuclear
shells, where most of the deformed and transitional nuclei are observed. It
is easy to check that the energies of the $0^{+}$\ excited states may be
described by simple phenomenological formula : 
\begin{equation}
E_{n}=An-Bn^{2}  \label{En}
\end{equation}
as a function of integer classification parameter $n.\;$ In (\ref{En}) $A$
and $B$ are fitting parameters, $n$ is an integer number corresponding to
each of the $0^{+}$\ excited state. We can define the physical meaning of
the integer classification quantum number $n$ , as the number of monopole
phonons that determine the collective structure of $0^{+}$\ excited state .

Let us consider the collective Hamiltonian:

\begin{equation}
H=\mathbf{\alpha }R_{+}^{j}R_{-}^{j}+\mathbf{\beta }R_{0}^{j}R_{0}^{j}+\frac{%
\mathbf{\beta }\Omega ^{j}}{2}R_{0}^{j}  \label{HR}
\end{equation}

\bigskip with interaction parameters $\alpha $ , $\beta $ and $\Omega ^{j}=%
\frac{2j+1}{2}.$

In (\ref{HR}) $H$ is written in terms of the monopole phonon operators $%
R_{+}^{j}$, $R_{-}^{j}$ and $R_{0}^{j}$ constructed from the fermion
creation and annihilation operators $\alpha _{jm}^{\dagger }$,$\;\alpha
_{jm} $, confined on a j-orbit with projections $m=-j,...,j$ as: 
\begin{equation}
\begin{array}{ll}
R_{+}^{j}= & {\frac{1}{2}}\sum\limits_{m}(-1)^{j-m}\alpha _{jm}^{\dagger
}\alpha _{j-m}^{\dagger }\;\text{; \ \ } 
\begin{array}{ll}
R_{-}^{j}= & {\frac{1}{2}}\sum\limits_{m}(-1)^{j-m}\alpha _{j-m}\alpha
_{jm}\;,
\end{array}
\\ 
& 
\begin{array}{l}
\\ 
\begin{array}{ll}
R_{0}^{j}= & {\frac{1}{4}}\sum\limits_{m}(\alpha _{jm}^{\dagger }\alpha
_{jm}-\alpha _{j-m}\alpha _{j-m}^{\dagger })\;,
\end{array}
\end{array}
\end{array}
\label{RR}
\end{equation}

The operators $R_{+}^{j}$, $R_{-}^{j}$ and $R_{0}^{j}$ satisfy the
commutation relations :

\begin{equation}
\left[ R_{0}^{j},R_{\pm }^{j}\right] =\pm R_{\pm }^{j};\ \ \ \ \ \ \ \left[
R_{+}^{j},R_{-}^{j}\right] =2R_{0}^{j}  \label{com}
\end{equation}

In order to simplify the notations further we will omit the indices$\;j$.

Let us now present this Hamiltonian in terms of \ ''ideal'' boson creation
and annihilation operators $\ \;\;\left[ b,b^{\dagger }\right] =1\ ;\;\;\;%
\left[ b,b\right] =\left[ b^{\dagger },b^{\dagger }\right] =0$ using the
Holstein-Primakoff transformation \cite{4} for the operators $R_{+\text{, }%
}R_{-}\;$and $R_{0}$ : 
\begin{equation}
\begin{array}{ccc}
R_{-}=\sqrt{2\Omega -b^{\dagger }b}\;b\text{; \ \ \ } & R_{+}=b^{\dagger }%
\sqrt{2\Omega -b^{\dagger }b}\text{; \ \ \ } & R_{0}=b^{+}b-\Omega
\end{array}
\label{HP}
\end{equation}

These transformations conserve the commutation relations between $R_{+}$, $%
R_{-}$ and $R_{0}\;\;$operators \ . Thus for the Hamiltonian (\ref{HRR}) in
terms of the new boson creation and annihilation operators ( ''ideal''
bosons ) $b^{+},\;b$\ we have: 
\begin{equation}
H=Ab^{\dagger }b-Bb^{\dagger }bb^{\dagger }b
\end{equation}

The coefficients in (\ref{Hb}) are related to the old ones as $A=\alpha
(2\Omega +1)-\beta \Omega ;\ \ \ \ \ \ \ B=\ \alpha -\beta .$

Therefore the energy of any monopole excited state\ \ $\left| n\right\rangle
=\frac{1}{\sqrt{n!}}(b^{+})^{n}\left| 0\right\rangle ,$ where $b\left|
0\right\rangle =0\;$ can be obtained as: \ 
\begin{equation}
\left\langle n\right| Ab^{\dagger }b-Bb^{\dagger }bb^{\dagger }b\left|
n\right\rangle -\left\langle 0\right| Ab^{\dagger }b-Bb^{\dagger
}bb^{\dagger }b\left| 0\right\rangle =E_{n}=An-Bn^{2}  \label{Enb}
\end{equation}

$\bigskip \ $The Hamiltonian (\ref{HRR}) creates the same energy spectrum as
the phenomenological formulae (\ref{En}) if the integer classification
parameter $n$ is interpreted as the number of ideal bosons .

\bigskip Now we can label every $K^{\pi }=0^{+}$\ state by an additional
characteristic $n$ - number of monopole bosons determining it's collective
structure. The parameters $\ \alpha $ and $\beta $ of (\ref{HR}) are
evaluated by fitting the experimental energies of the different $0^{+}$
states of a given nucleus to the theoretical ones for all possible
distributions of the classification numbers $n.$ Some of our calculations
corresponding to the distributions with minimal value of $\chi $-square
along with the experimental data are presented on \textbf{Figure 1.}

There is a relatively good agreement with experiment even in the case of $%
^{194}$Pt having eight $0^{+}$ states. It is very important to point out
that the ordering of the states in respect to their number of phonons does
not correspond to increase of energy. For most of the nuclei the lowest
excited $K^{\pi }=0^{+}$ states have more collective nature ( lager $n$ )
than the states with higher energies. The values of the parameters $\alpha $
and $\beta $ for some of the considered nuclei are given in \textbf{Table 1}%
. 
\begin{equation}
\tag*{\QTR{bf}{Table
1.}
Values
of
hte
parameters
$a$
and
$\beta
$\
for
some
nuclei. \
\ \
\ \ \ \
\ \
\ \ \
\ \ }
\end{equation}

\bigskip\ \ \ \ \ \ \ \ \ \ 
\begin{tabular}{|l|l|l|l|l|}
\hline
Nucleus & $^{114}$Cd & $^{156}$Gd & $^{158}$Er & $^{168}$Yb \\ \hline
$a$ & 1.19593 & 1.17889 & 1.23148 & 1.17889 \\ \hline
$\beta $\  & 0.28923 & 0.397842 & 0.21933 & 0.397842 \\ \hline
Nucleus & $^{172}$Yb & $^{178}$Hf & $^{188}$Os & $^{194}$Pt \\ \hline
$a$ & 1.3206 & 1.21661 & 0.762154 & 0.763518 \\ \hline
$\beta $\  & 0.35529 & 0.318346 & 0.101847 & \ 0.072 \\ \hline
\end{tabular}
\smallskip \bigskip

The next step is the investigation of the rotational ''$\beta $-bands''
build on the already considered as band heads $0^{+}$states. We use the
standard quantum rigid rotator approach, in which energies of the states
belonging to a given rotational band : 
\begin{equation}
E_{rot}(L)=\frac{L(L+1)}{2\mathcal{F}}  \label{Erot}
\end{equation}
are determined by the moment of inertia $\mathcal{F}$ . In order to obtain
it, we restrict ourselves to nuclei with axial symmetry and uniform density
distribution. In our approach we use the Kassini form for the description of
the nuclear surface. In the $YZ$ plane this form is expressed as: 
\begin{equation*}
Y^{\;2}=\sqrt{c^{4}+4a^{2}Z^{2}}-a^{2}-Z^{2}
\end{equation*}
and is plotted on \textbf{Figure 2.\ } The radius of the sphere inside the
plain cross section $R_{sph}=\sqrt{a^{2}-c^{2}}$ and the maximal distance
from the center $R_{\max }=\sqrt{a^{2}+c^{2}}$ are conveniently expressed by
means of the parameters $a$ and $c$ and illustrate their meaning.

\bigskip The choice of the Kassini form, which depends on only two
parameters $a$ and $c$, is determined by its simplicity and the possibility
to get exact, analytical expressions for many physical observables depending
on the nuclear shape. For instance the main component of the tensor of
inertia in the case of rotation around fixed $Y$ axis ({\small see}\ {\small %
Figure 2}) is:

\begin{eqnarray}
\mathcal{F}_{y} &=&m\rho _{0}\frac{\pi \,\left( 2\,\,\sqrt{a^{2}+c^{2}}%
\,\left( 16\,a^{6}-8\,a^{4}\,c^{2}+246\,a^{2}\,c^{4}+15\,c^{6}\right)
\right) }{960\,a^{2}}+  \notag \\
&&  \label{Min} \\
&&m\rho _{0}\frac{\pi \left( 15\,c^{8}\,\log (\frac{c^{2}}{%
2\,a^{2}+c^{2}+2\,a\,\sqrt{a^{2}+c^{2}}})-240\,a^{4}\,c^{4}\,\log (\frac{%
2\,a^{2}+c^{2}+2\,a\,\sqrt{a^{2}+c^{2}}}{c^{2}})\right) }{960\,a^{3}}  \notag
\end{eqnarray}

In the expression (\ref{Min}) $m$ is the mass of the nucleons, $A$ is a
nucleus mass number and $\rho _{0}=\frac{A}{V(a,c)}$ is the value of the
matter density in the nucleus , where $V(a,c)$ is its volume: \ 
\begin{equation}
V(a,c)=\frac{2R_{\max }\,\left( R_{sph}^{2}-a^{2}\right) \,\pi }{6}+\frac{%
c^{4}\,\pi \,ArcSinh(\frac{2\,a}{c^{2}\,\sqrt{\frac{1}{a^{2}+c^{2}}}})}{2\,a}
\label{Obem}
\end{equation}
Another important quantity is the mean square radius, which is obtained to
be:

\begin{equation}
R_{ms}^{2}(a,c)=\frac{a\,\left(
4\,a^{4}+24\,c^{4}-2\,a^{2}\,c^{2}-15\,a\,c^{4}\,\sqrt{\frac{1}{a^{2}+c^{2}}}%
\,\log (\frac{c^{2}+2\,a\,\left( a+\sqrt{a^{2}+c^{2}}\right) }{c^{2}}%
)\right) }{\frac{3\,c^{4}\,ArcSinh(\frac{2\,a\,\sqrt{a^{2}+c^{2}}}{c^{2}})}{%
\sqrt{a^{2}+c^{2}}}-20\,a^{3}+10\,a\,c^{2}}  \label{Rms(a,c)}
\end{equation}

\bigskip In the realistic applications to the nucleus we are going to use
instead of the component $\mathcal{F}_{y}$ of the tensor of inertia the
difference $\mathcal{F=F}_{y}-\mathcal{F}_{sph}$, where the moment of
inertia $\mathcal{F}_{sph}$\ corresponds to the sphere with radius $R_{sph}=%
\sqrt{a^{2}-c^{2}}$(see \textbf{Figure 2}) and is given by: 
\begin{equation}
\mathcal{F}_{sph}\ =m\rho _{0}\frac{8\,{\left( c^{2}-a^{2}\right) }^{\frac{5%
}{2}}\,\pi }{15}  \label{Insph}
\end{equation}

As a result in the case of nuclei with spherical symmetry we directly come
to $\mathcal{F}=0$ because of the condition $a=0$ for the Kassini form. The
use of the difference $\mathcal{F}$ \ is further approved by the realistic
values of mean square radii without the necessity to involve additional
parameters. Among the advantages of the Kassini form one can point out that
it permits easy introduction of different proton and neutron density
distributions, which gives possibility to introduce other types of
collective motion, for instance ''scissors-type'' modes etc.

Now we will make use of the quantity $\Delta A$ which is the difference
between the particles in the whole nuclear volume and the ones in the sphere
inscribed in the middle of the Kassini form, which we consider as the number
of nucleons responsible for the rotational motion: 
\begin{equation}
\Delta A=\frac{\rho _{0}}{6}\pi \,\left( \sqrt{{\small a}^{2}{\small +c}^{2}}%
\,\left( {\small 2\,c}^{2}{\small -4\,a}^{2}\right) +\,\frac{{\small 3\,c}%
^{4}}{{\small a}}ArcSinh{\Large (}\frac{{\small 2\,a}}{{\small c}^{2}{\small %
\,}\sqrt{\frac{1}{a^{2}+c^{2}}}}{\Large )}-{\small 8}\,{\left( {\small c}^{2}%
{\small -a}^{2}\right) }^{\frac{3}{2}}\right)  \label{DelA}
\end{equation}

Finally, to link our moment of inertia $\mathcal{F=F}_{y}-\mathcal{F}_{sph}$
with the number of bosons of any $K^{\pi }=0_{n}^{+}$\ $\;$state we imply
the results of the Nuclear Surface Oscillation Model \cite{3} and apply the
recently derived \cite{4} expressions for the commutation relations :

\begin{equation}
\left[ b^{n},(b^{\dagger })^{m}\right] =\left\{ 
\begin{array}{c}
\sum\limits_{l=0}^{n-1}\frac{m!}{(m-n+l)!}\binom{n}{l}(b^{\dagger
})^{m-n+l}b^{l}\;\;\;\;\;\;\;\;\;n\leqslant m\  \\ 
\\ 
\sum\limits_{l=0}^{m-1}\frac{n!}{(n-m+l)!}\binom{m}{l}(b^{\dagger
})^{l}b^{n-m+l}\;\;\;\;\;\;\;\;\;\;\;\;n\geqslant m\ 
\end{array}
\right.  \label{[bb]}
\end{equation}

The following expression of the \ mean square radius of spherical nucleus
with uniform density, in a collective excited state $\left| n\right\rangle :$

\begin{eqnarray}
R_{ms}^{2}(n) &=&\frac{3\,{c}^{2}\,\left( 32\,{\pi }^{2}+80\,n\,\pi \frac{%
\,E_{0}}{C_{0}}+15\,\left( -1+n\right) \,n(\,\frac{\,E_{0}}{C_{0}}%
)^{2}\right) }{160\,{\pi }^{2}+60\,n\,\frac{\,E_{0}}{C_{0}}}  \notag \\
&&  \label{Rms1} \\
\text{if }n &=&0\text{ we have \ }R_{ms}^{2}(0)=\frac{3\,{c}^{2}\,}{5} 
\notag
\end{eqnarray}
is obtained in the Nuclear Surface Oscillation Model \cite{3}. \bigskip $%
E_{0}$ is the one-phonon excitation energy, $C_{0}$ - nuclear surface
compressibility parameter and $c$ is the radius at half density .

Taking into account that the ratio of $\frac{\,E_{0}}{C_{0}}\ll 1$ and
imposing the restriction, that the dependence of the mean square radius on
the number of bosons in our case (\ref{Rms(a,c)}) and the spherical case $%
R_{ms}^{2}(n)$ (\ref{Rms1})\ are the same , after some algebraic
calculations we obtain the dependence of the parameter $a$ on $n$ . \ The
parameter $a$ separated in two terms: 
\begin{equation*}
a(n)\thickapprox a+\Delta a(n),
\end{equation*}
where the part directly related to the collective motion is:\bigskip\ \ 
\begin{equation}
\begin{array}{c}
\Delta a(n)=\frac{\sqrt{\frac{3}{5}}\,\sqrt{c^{2}\,n\,\frac{E{_{0}}\,}{{{%
C_{0}}}}\left( 8\,{\pi }^{2}\,\left( -3+20\,\pi \right) \,+3\,\left( -10\,{%
\pi }^{2}+n\,\left( 3-20\,\pi +10\,{\pi }^{2}\right) \right) \frac{E{_{0}}}{{%
{C_{0}}}}\,\right) }}{8\,{\pi }^{2}}
\end{array}
\label{a(n)}
\end{equation}
\smallskip

As a result, the moment of inertia \ $\mathcal{F=F}_{y}-\mathcal{F}_{sph}$
becomes the following function of $n$ - the number of phonons : 
\begin{equation}
\mathcal{F}(a,c)=\mathcal{F}_{y}[a{\small +}\Delta a(n),c]-\mathcal{F}%
_{sph}[a+\Delta a(n),c]  \label{IN(n)}
\end{equation}
We apply this expression (\ref{IN(n)}) in our calculations of realistic
nuclear systems, at the same time fixing the value of the parameter $c$ $%
=1.286A^{\frac{1}{3}}$ , where $A$ is the mass number . In \textbf{Figure 3}
we present the dependence of $\Delta A$ (\ref{DelA}), the moments of inertia 
$\mathcal{F}_{y}$ and $\mathcal{F}(a,c)$ (multiplied by 10$^{-3}$) and the
mean square radius on the parameters of the Kassini form . $\Delta A$ in
each case is less than the number of valent nucleons ( some indication of
the role of super-conductivity ). $\mathcal{F}(a,c)$ increases with $a$ to
it's limiting value, corresponding to the rigid moment of inertia $\mathcal{F%
}(c,c)$. For instance in the case of $^{174}$Hf \ $\mathcal{F}(c,c)$
=13489.5 Mev fm$^{2}.$

Now we have all the necessary tools to calculate the energies of rotational
bands. Using formula (\ref{En}) we first analyze the distribution of $0^{+}$%
\ states for the nuclei under consideration. Then with the fixed number of
phonons $n$ for each band head , we fit the model parameter $a$ entering in
the moment of inertia (\ref{IN(n)}) from the experimental ground state ($n=0$%
) rotational band using (\ref{Erot}). The second parameter $\frac{E{_{0}}}{{{%
C_{0}}}}$ can be determined from any rotational band built on excited $0^{+}$%
\ state. We choose the one built on the lowest in energy of the excited band
heads. There is nothing amazing in the fact that using the crossing of
rotational bands we could successfully describe the energy behavior of the
yrast lines in the cases $\ $of $^{174}$Hf, $^{168}$Yb. It is surprising,
although that we have extremely good agreement with experiment using only
two fitting parameters $a$ and $\frac{E{_{0}}}{{{C_{0}}}}$, and applying
simple ''rigid rotator'' model (\ref{Erot}). Our results \ (shown with lines
) along with the experimental data and the values of the model parameters
are presented on \textbf{Figures 4 and 5}. On \textbf{Figure 6. }we present
the results of our calculations of rotational bands for the nucleus $^{168}$%
Yb in the case when $n=0$ for all the bands heads. This example illustrates\
the importance of including the bands heads' structure in the calculation of
the spectrum of the rotational bands.

We introduce the quantity $\mathbf{\beta }_{K}\mathbf{=}$ $\frac{R_{\max
}-R_{sph}}{R_{sph}},$ which we consider as a measure of the deformation of
the nuclear system. As mentioned above the values of the mean square radii $%
R_{ms}\;$and also the ''deformations'' \ $\mathbf{\beta }_{K}$ \ are
relatively realistic and in very good agreement with the corresponding
values obtained in the HFB approach \cite{5}. Comparison of the \
corresponding values are shown in \textbf{Table 2. }

\bigskip \textbf{Table 2.} Values of model parameters and comparison with
HFB data \cite{5}. 
\begin{equation*}
\begin{tabular}{|l|l|l|}
\hline
& $^{168}Yb$ & $^{174}Hf$ \\ \hline
$a\;$ & $3.65$ & $3.50$ \\ \hline
$R_{ms}\;$ & $5.31$ & $5.39$ \\ \hline
$^{(HFB)}R_{ms}\;$ & $5.298$ & $5.365$ \\ \hline
$\mathbf{\beta }_{K}$ & $0.31$ & $0.27$ \\ \hline
$^{(HFB)}\beta $ & $0.317$ & $0.317$ \\ \hline
$\frac{E_{0}}{C_{0}}$ & $0.015$ & $0.013$ \\ \hline
\end{tabular}
\end{equation*}

\bigskip It is important to point out the particular behavior \ of \ the
mean square radius, as a function of $a$ (see \textbf{Figure 3} ). It has a
minimum in the region close to the estimated values of this parameter . In
the future, it is interesting to investigate this phenomena from physical
point of view.

There is a small difference between our calculations of the energies and
their experimental values at the angular momenta, where the ''back bending''
may occur. This is more explicit on \textbf{Figure 7, } where we present the
behavior of the first derivatives of the calculated and experimental
yrast-lines energies , in respect to the angular momentum $L.$ The
explanation of the mechanism of phase transition, caused by the jump from
the ground band to another bands that may take place is a very interesting
problem, but was not exactly the aim of this paper. We have prepared a
reasonable background for such an investigations.

The presented results are rather convincing for the possibility of this
phenomenological approach, assuming the Kassini form for the nuclear system,
to describe the rotational bands built on the 0$^{+}$ states with the input
information about their structure as the band heads. Even more the structure
of the yrast band is accurately reproduced by the behavior of these
rotational bands. On the other hand , applying this model to the analysis of
the yrast lines one can get the direct information about the collective
structure of the 0$^{+}$ states that appear as band heads of corresponding
crossing rotational bands, which form it. Also, applying our results for the
yrast band to the traditional variable moment of inertia approaches $%
\mathcal{F}(L),$\ one can find more reasonable physical meanings of their
model parameters, together with the explanation of the behavior of energies
in the regions of crossing of the different rotational bands.

I thank sincerely A. Georgieva, V. Zamfir , C. Giusti and M. Soitsov for
fruitful discussions and help.

This work was partially supported by the Bulgarian Science Committee under
contract number $\Phi $ 905.

\bigskip 
\begin{equation}
\FRAME{itbpF}{3.7931in}{5.239in}{0in}{}{}{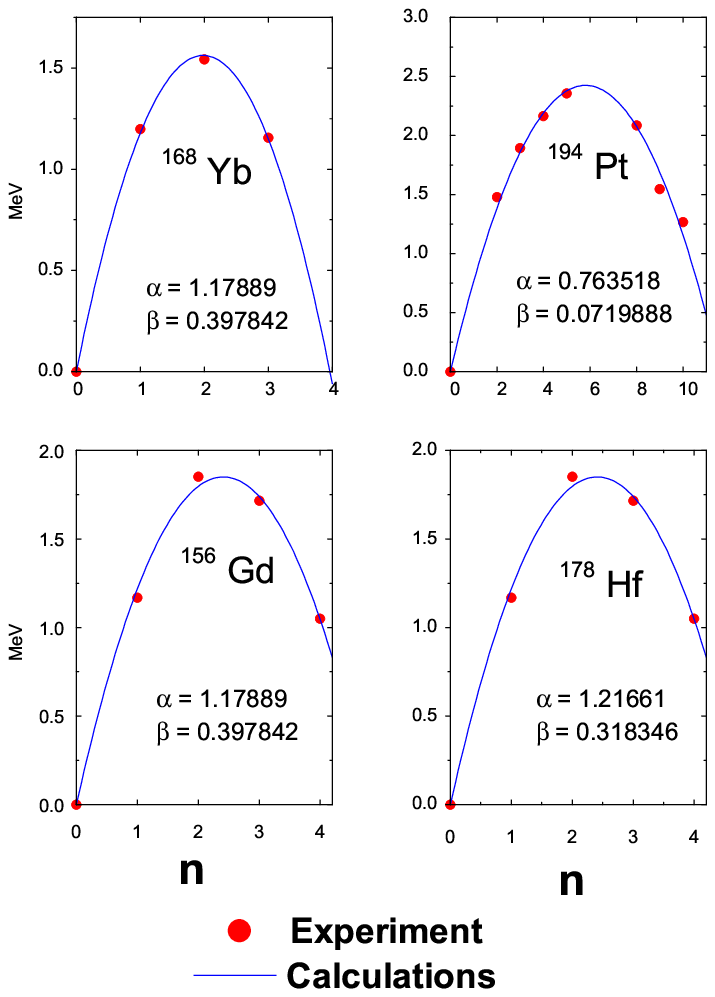}{\special{language
"Scientific Word";type "GRAPHIC";display "USEDEF";valid_file "F";width
3.7931in;height 5.239in;depth 0in;original-width 3.1332in;original-height
3.9297in;cropleft "0";croptop "1";cropright "1";cropbottom "0";filename
'allnuli1.eps';file-properties "XNPEU";}} 
\tag*{Figute
1.
\QTR{small}{Number of
phonon
distributions
of
}$0^{+}$\QTR{small}{
state
energies for
different
isotopes.}}
\end{equation}

\bigskip 
\begin{equation}
\FRAME{itbpF}{3.0978in}{2.6221in}{0.0104in}{}{}{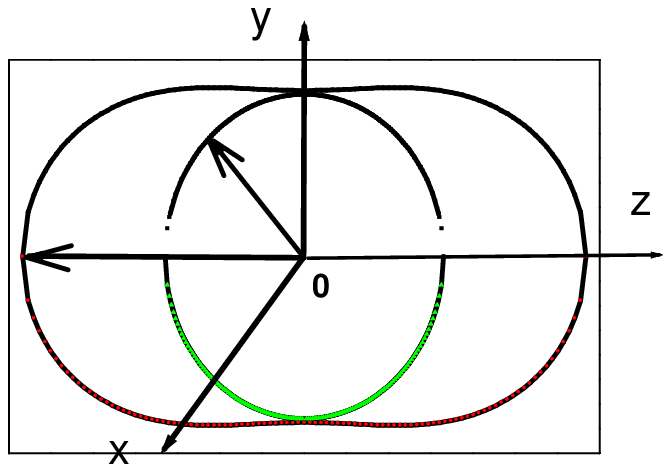}{\special%
{language "Scientific Word";type "GRAPHIC";display "USEDEF";valid_file
"F";width 3.0978in;height 2.6221in;depth 0.0104in;original-width
7.8672in;original-height 11.336in;cropleft "0";croptop "0.9320";cropright
"0.4351";cropbottom "0.5629";filename 'kassini.eps';file-properties "XNPEU";}%
}  \tag*{Figure 2. Kassini form in YZ plane. \ \ \ \ \ \ \ \ \ \ }
\end{equation}
\begin{equation}
\FRAME{itbpFw}{3.9678in}{4.6726in}{0in}{}{}{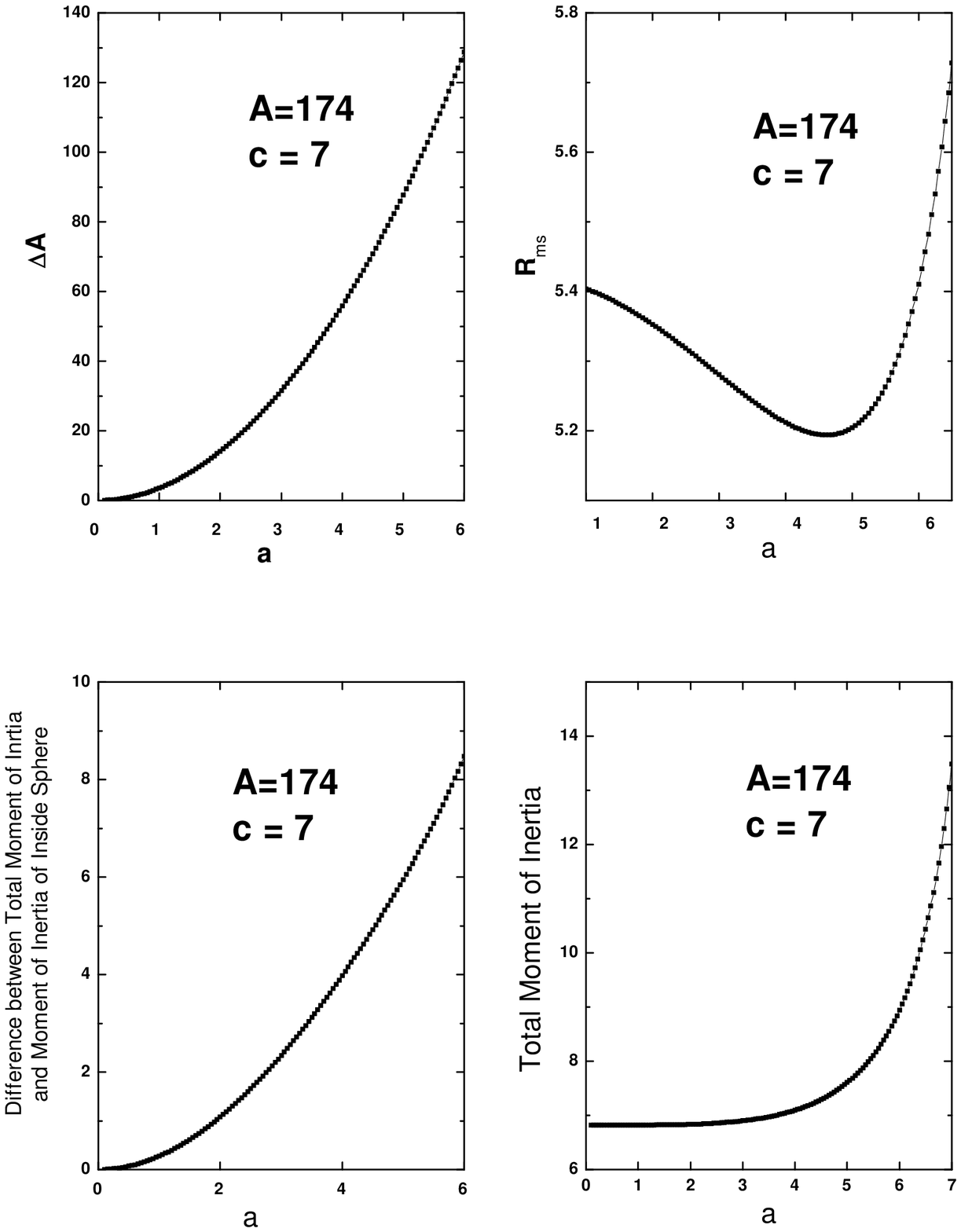}{\special{language
"Scientific Word";type "GRAPHIC";display "USEDEF";valid_file "F";width
3.9678in;height 4.6726in;depth 0in;original-width 7.8793in;original-height
11.2962in;cropleft "0";croptop "1";cropright "1";cropbottom
"0.1777";filename 'arin.eps';file-properties "XNPEU";}}  \tag*{Figure
3.\QTR{small}{ Dependence of mean square radius and moment of inertia on
Kassini form parameters}}
\end{equation}

\bigskip 
\begin{equation}
\FRAME{itbpF}{4.4659in}{5.8816in}{0in}{}{}{ybpaper.eps}{\special{language
"Scientific Word";type "GRAPHIC";display "USEDEF";valid_file "F";width
4.4659in;height 5.8816in;depth 0in;original-width 3.0398in;original-height
3.614in;cropleft "0";croptop "1";cropright "1";cropbottom "0";filename
'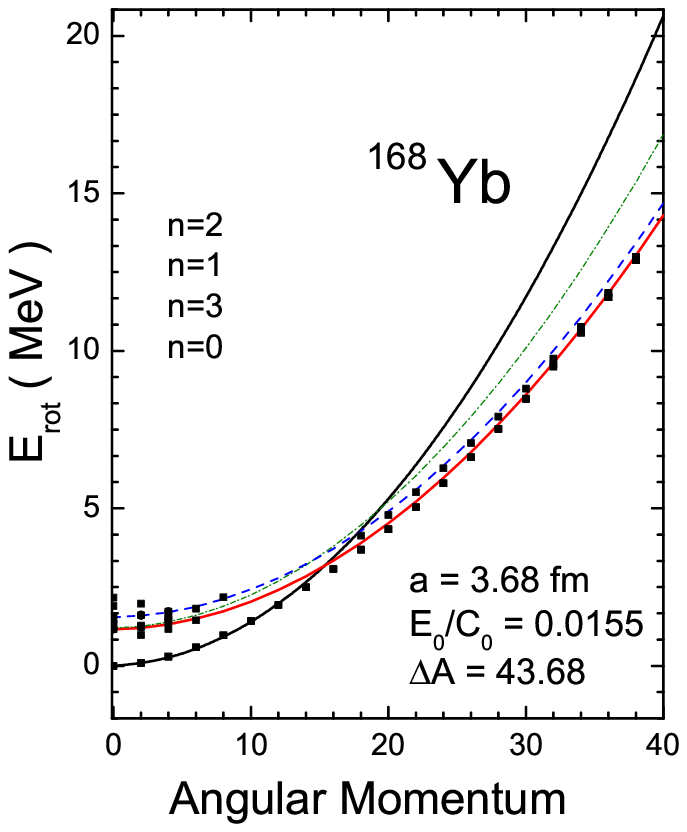';file-properties "XNPEU";}} 
\tag*{Figure
4.%
\QTR{small}{
Comparison
of
calculated
rotational
energies
with
experiment
for
}$^{168}$%
\QTR{small}{Yb. \ }}
\end{equation}

\bigskip 
\begin{equation}
\FRAME{itbpF}{4.1684in}{5.8703in}{0in}{}{}{hfpaper.eps}{\special{language
"Scientific Word";type "GRAPHIC";display "USEDEF";valid_file "F";width
4.1684in;height 5.8703in;depth 0in;original-width 2.2338in;original-height
3.1557in;cropleft "0";croptop "1";cropright "1";cropbottom "0";filename
'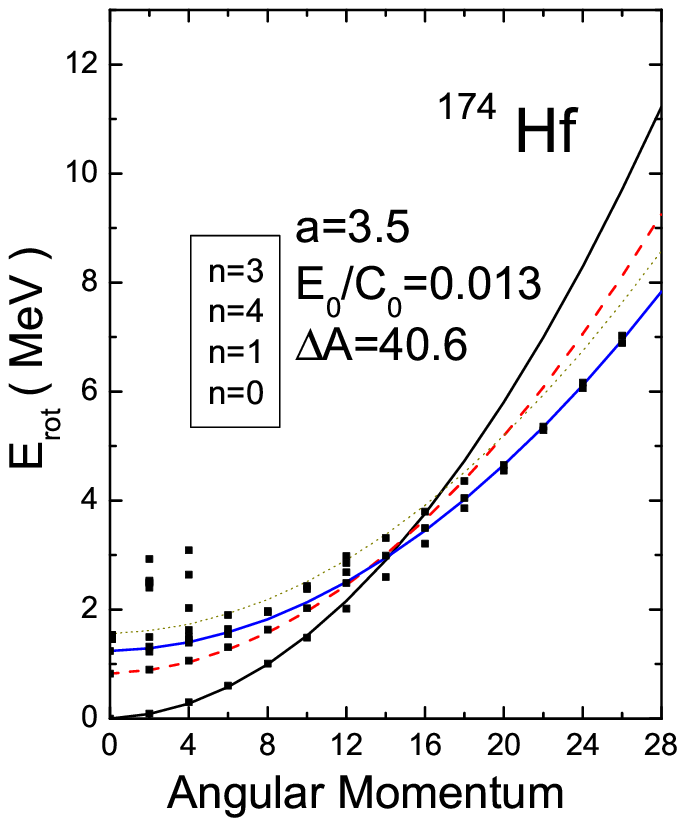';file-properties "XNPEU";}} 
\tag*{Figure
5.%
\QTR{small}{
Comparison
of
calculated
rotational
energies
with
experiment
for
}$^{174}$%
\QTR{small}{Hf. \ }}
\end{equation}

\bigskip 
\begin{equation}
\FRAME{itbpF}{4.3543in}{5.5651in}{0in}{}{}{yb168n0.eps}{\special{language
"Scientific Word";type "GRAPHIC";display "USEDEF";valid_file "F";width
4.3543in;height 5.5651in;depth 0in;original-width 3.1808in;original-height
3.4186in;cropleft "0";croptop "1";cropright "1";cropbottom "0";filename
'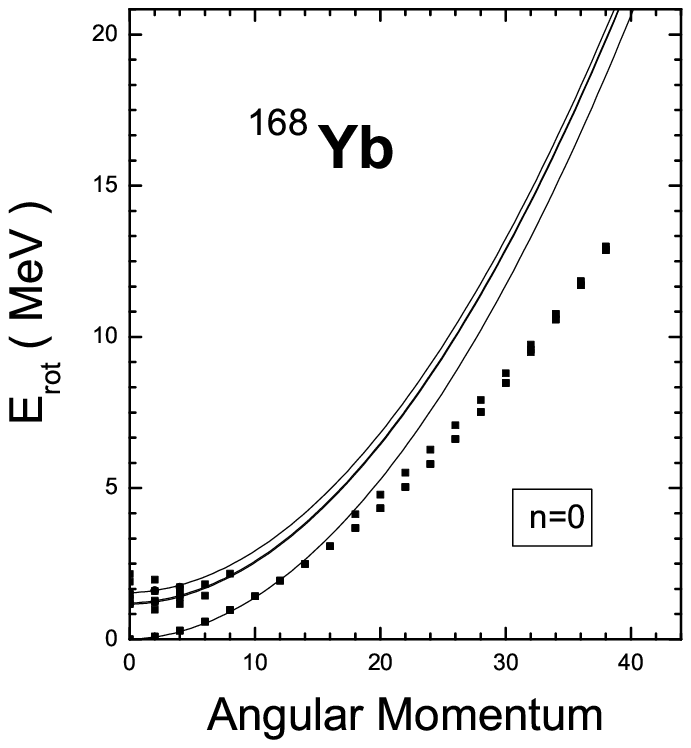';file-properties "XNPEU";}} 
\tag*{Figure
6.
\QTR{small}{Results
in
the
case
when
\smallskip%
\
moment
of
inertia
does
not
depend
on
\QTR{large}{n}
.}}
\end{equation}

\begin{equation}
\FRAME{itbpFw}{5.38in}{6.877in}{0in}{}{}{deriv.eps}{\special{language
"Scientific Word";type "GRAPHIC";display "USEDEF";valid_file "F";width
5.38in;height 6.877in;depth 0in;original-width 2.9179in;original-height
2.3436in;cropleft "0.0935";croptop "1.0279";cropright "1.0977";cropbottom
"0";filename '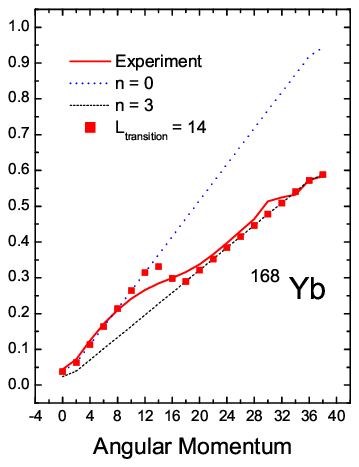';file-properties "XNPEU";}} 
\tag*{Figure
7.
First
derivatieves
by
L
for
experimental,
n=0
and
n=3
rotational bands.}
\end{equation}

\bigskip

\bigskip

$\bigskip $


\begin{thebibliography}{9}
\bibitem{1}  S. T. Belyaev \ Mat. Fys. Medd \ Dan. Vid Seelgk.31, {\small N}%
11,(1959); Wenes G. et al. - Phys. Rev. C23, 1981, p.2291; \ Van Hieven J.
F. A. et al. - Nucl. Phys. A269, 1976, p.159; Sakahura M. et al.\ - Z.
Physik A289,\ 1979, \ p. 163 ; Shikata Y. et al. - Z. Physik A300, 1981, p.
217;\ Lopak V and V.Paar - Nucl.Phys. A297, 1978, p.471; De Vries H. F., P.
J. Brussard\ \ - Z. Physik A286, 1978, p.1; Ahalpara D. P. et al. - Nucl.
Phys. A371, 1981, p.210;\ Arima A., F. Iachello - Ann. Phys 99,1976. p, 253;
111, 1978, p. 201; 123, 1979, p. 468; \ Duval P. D. B. R. Barret \ - Phys.
lett. 100B, 1981, p.223 and Nucl. Phys. A376, 1982, p. 213; \ Druce C.H. et
al. - Nucl. Phys. 8, 1982, p.1565; \ V. P. Garistov Bulg. J. Phys.14,
(1987), 4, 317; Tazaki S. et al. - Prog. Theor. Phys. 71,1981, ch.4; Cohen
T. D. - Nucl. Phys. A436, 1985, p. 16; C. Volpe et al. Nucl. Phys A 647
(1999),246; A. K. Kerman Annals of Physics 12, (1961), 300; D. M. Brink,
A.F.R. De Toledo Piza, A.K. Kerman - Phys. Lett. V.19, \#5, (1965), 413,
Kishimoto T., T. Tamura - Nucl. Phys. A192, 1972, p.246;D.R.Bes and R.A.
Sorensen - Adv. in Nucl. Phys. 2 (1969), p. 129; \ \ Julin R. et al. - Z.
Physik A296, 1980, p. 315; Mheemeed A. et al. - Nucl. Phys. A412, 1984,
p.113; A. Aprahamian et al. - Phys. Lett. 140B,1984, 1-2, p.22; \ X. Wu, A.
Aprahamian, J. Castro-Ceron- Phys. Lett. 316B,1983 ; \ Aprahamian A. et al.
- Phys. Rev. C,49,\#4 (1994) ; Kantele J. et al. - Z. Physik A289, 1979,
p.157;\ J. A. Cizewski et al. - Nucl. Phys. A323 (1979),349; W.
Andrejtscheff et al. Nucl. Phys. 12 (1986)L151.

Mitsuo Sacai Atomic Data and Nuclear Data Tables 31, 399-432 (1984); A.M.I.
Haque et al. Nucl. Phys. A 455, 231 ( 1986 ); R. S. Hager and E. C. Seltzer
NDT A4, 1 (1968 )

\bibitem{2}  \ T.Holstein, H.Primakoff, Phys. Rev. \textbf{58}, (1940) 1098;

A.O. Barut, Phys. Rev. \textbf{139}, (1965) 1433;\textbf{\ }

R. Marshalek Phys. Lett. \textbf{B }97 (1980) 337;

C. C. Gerry, J. Phys. \textbf{A} 16, (1983) 11.

\bibitem{3}  A. N. Antonov, V. P. Garistov, I. J. Petkov Phys. Lett. B; V.
P. Garistov \ Bulg. J. Phys. 14 (1987), 4, 317; Vladimir P. Garistov IJMP 
\textbf{E }v.4 \#2 (1995),371

\bibitem{4}  V P.Garistov, P. Terziev, preprint nucl-th/9811100 (1998)\ 

\bibitem{5}  M. Stoitsov private communications.
\end{thebibliography}
\end{document}